\documentclass[runningheads]{llncs}
\usepackage{graphicx}
\usepackage{todonotes}
\usepackage{subfiles}
\usepackage{caption}
\usepackage{subcaption}
\usepackage{hyperref}
\hypersetup{
    colorlinks=true,
    linkcolor=blue,
    filecolor=magenta,      
    urlcolor=cyan,
}
\begin{document}
%
\title{Surface Agnostic Metrics for Cortical Volume Segmentation and Regression}
\author{Samuel Budd\inst{1}\orcidID{0000-0002-9062-0013},
Prachi Patkee\inst{2},
Ana Baburamani\inst{2},
Mary Rutherford\inst{2},
Emma C. Robinson\inst{2}, 
Bernhard Kainz\inst{1}}
\authorrunning{S. Budd et al.}
\institute{Imperial College London, Dept. Computing, BioMedIA, London, UK  \and
Centre for the Developing Brain, School of Biomedical Engineering and Imaging Sciences, King's College London, ISBE, London, UK \\ \email{samuel.budd13@imperial.ac.uk}}
\maketitle              
\begin{abstract}
The cerebral cortex performs higher-order brain functions and is thus implicated in a range of cognitive disorders. Current analysis of cortical variation is typically performed by fitting surface mesh models to inner and outer cortical boundaries and investigating metrics such as surface area and cortical curvature or thickness. These, however, take a long time to run, and are sensitive to motion and image and surface resolution, which can prohibit their use in clinical settings. In this paper, we instead propose a machine learning solution, training a novel architecture to predict cortical thickness and curvature metrics from T2 MRI images, while additionally returning metrics of prediction uncertainty. Our proposed model is tested on a clinical cohort (Down Syndrome) for which surface-based modelling often fails. Results suggest that deep convolutional neural networks are a viable option to predict cortical metrics across a range of brain development stages and pathologies.
\end{abstract}

\section{Introduction}
 Irregularities in cortical folding and micro-structure development have been implicated in a range of neurological and psychiatric disorders including: Autism, where disruptions to folding cortical and thinning of the cortex has been found in regions associated with social perception, language, self-reference and action observation~\cite{Yang2016CorticalGyrification}; Down Syndrome, where smoother cortical surfaces and abnormal patterns of cortical thickness are linked to impaired cognition~\cite{Lee2016DissociationsThickness.}; Epilepsy, where malformations in cortical development are associated with seizure onset~\cite{Leventer2008MalformationsEpilepsy} and psychosis, which is associated with abnormal functional behaviour of the pre-frontal cortex~\cite{Mukherjee2016DisconnectionDisorders.}.

There is a strong need to model cortical development in at-risk neonatal populations. However, due to the heterogeneous and highly convoluted shape of the cortex, it has proved highly challenging to compare across populations. Recent consensus has been that cortical features are best studied using surface-mesh models of the brain~\cite{Glasser2016ACortex}, as these better represent the true geodesic distances between features on the cortex. However, these require running of costly multi-process pipelines which perform intensity-based tissue segmentation, followed by mesh tessellation and refinement. 

For developmental cohorts, the fitting of surface mesh models is even more challenging due to the relatively low resolution and likely motion corruption of these datasets. This leads to artifacts and partial volume effects or blurring across tissue boundaries. Methods to tackle these problems individually exist~\cite{Makropoulos2018TheReconstruction} but they are highly tuned to high-resolution, low motion research data sets and do not always transfer well to clinical populations as outlined in Figure~\ref{fig:sota}. 

As an alternative, several groups have proposed techniques for extracting cortical surface metrics from volume data directly~\cite{Dahnke2013CorticalEstimation,Tustison2014Large-scaleMeasurements,Das2009RegistrationMeasurement}. Specifically,
Tustison et al~\cite{Tustison2014Large-scaleMeasurements,Das2009RegistrationMeasurement} show that an ANTs-based extension for the estimation of cortical thickness from volumetric segmentation generates thickness measures which outperform FreeSurfer (surface) metrics when applied to predictive tasks associating thickness with well-studied phenotype relations such as age and gender. Nevertheless, volumetric fitting approaches such as~\cite{Tustison2014Large-scaleMeasurements} have not yet been validated on developmental data and their slow run times limit their utility for clinical applications. In this paper we therefore seek to develop a novel algorithm for cortical metric prediction from clinical, developmental data, through Deep Learning. 
\noindent\textbf{Contribution:}
The key contributions of this paper are: 1) We propose the first probabilistic Deep Learning tool for cortical segmentation and metric learning. 2) We validate the method against cortical thickness and curvature prediction for data from the Developing Human Connectome Project (dHCP); 3)  The tool is used to predict cortical metrics for a Down Syndrome cohort in which surface-based analysis often fails; 4) Our probabilistic approach returns confidence maps which can inform clinical researchers of areas of the brain where measurements are less reliable. 

\section{Related work}

The cerebral cortex is a thin layer of grey-matter tissue at the outer layer of the brain. Studying it is important for improved understanding of cognitive and neurological disorders but doing so is challenging due to it's complex shape and patterns of micro-structural organisation.

\begin{figure}[h!]
    \centering
    \includegraphics[width=0.8\textwidth]{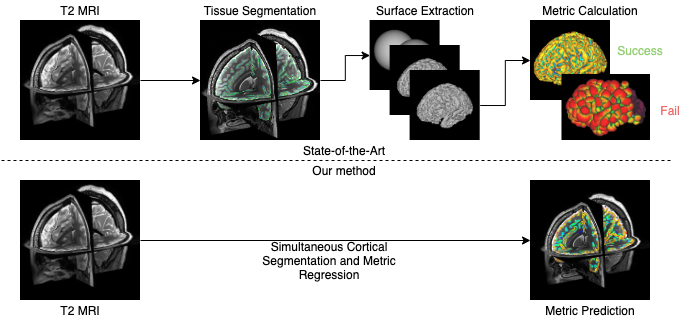}
    \caption{Overview figure of current state-of-the-art approach to extracting cortical metrics from MRI volumes showing a success and failure case vs. our method. The segmentation and surface extraction components of the pipeline are prone to fail, thus to produce artifacts as displayed in the fail case above.}
    \label{fig:sota}
\end{figure}

Currently, most studies of the cortex use surface mesh models~\cite{Makropoulos2018TheReconstruction,Glasser2013TheProject,Fischl2000MeasuringImages}, which fit mesh models to inner and outer cortical boundaries following pipelines which perform tissue segmentation, followed by surface tessellation with intensity based refinement (Figure \ref{fig:sota}). Summary measures of cortical thickness and curvature may then be estimated from the Euclidean distance between mesh surfaces, and principal curvatures of the white-matter surface respectively.

By contrast, ANTs (Advanced Normalisation Tools) and CAT (Computational Anatomy Toolbox) propose volume-based pipelines for cortical thickness estimation. Specifically, the ANTs  pipeline estimates cortical thickness in five steps: 1) Initial N4 bias correction of input MRI; 2) Segmentation/Registration based brain extraction; 3) Alternating prior-based segmentation and weighted bias correction using Atropos and N4; 4) DiReCT-based cortical thickness estimation; 5) Optional normalisation to template and multi-atlas cortical parcellation~\cite{Tustison2014Large-scaleMeasurements,Das2009RegistrationMeasurement}. CAT uses segmentation to estimate white matter (WM) distance, and then projects the local maxima (equal to cortical thickness) to other grey matter voxels by using a neighbour relationship defined by the WM distance~\cite{Dahnke2013CorticalEstimation}.
\section{Method}

\begin{figure}[h!]
    \centering
    \includegraphics[width=0.8\textwidth]{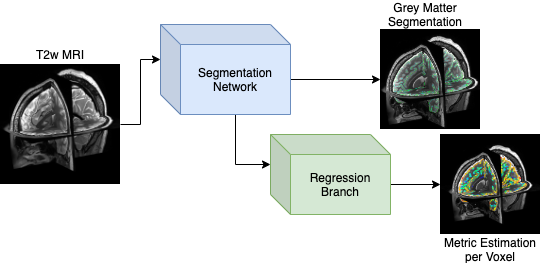}
    \caption{Architecture Diagram for the proposed simultaneous segmentation and regression network: The network predicts a cortical segmentation through the chosen segmentation architecture while simultaneously regressing a cortical metric value for every voxel in the image via the fully connected regression branch}
    \label{fig:regnet}
\end{figure}

\noindent\textbf{Segmentation and Regression Network}:
Our proposed network architecture augments the popular U-Net architecture into a 3D Multi-Task prediction network~\cite{RonnebergerU-Net:Segmentation}. We introduce a branch of fully connected layers prior to the final convolution of the U-Net as shown in Figure \ref{fig:regnet}. The network predicts a cortical segmentation through the standard U-Net architecture while simultaneously regressing a cortical metric value for every voxel in the image~\cite{Cicek20163DAnnotation}. These two tasks are strongly coupled and as such we design the regression branch of our network to see a large amount of information from the segmentation path. We use a cross-entropy loss function for the segmentation task and consider two different loss functions for the regression task: Mean squared error (MSE) and Huber/L1 Loss, the latter is considered to encourage smoothness of regression predictions across neighbouring pixels by being robust to outliers~\cite{Huber1964RobustParameter}, a property that should hold for cortical metric predictions. 

We propose a probabilistic extension of our network in which we introduce Dropblock to our network during training and test time, this approach ensures variation in our network predictions during inference, this forms the baseline for our probabilistic experiments~\cite{Ghiasi2018DropBlock:Networks}. We propose an alternative probabilistic segmentation and regression architecture based on the PHiSeg network~\cite{Baumgartner2019PHiSeg:Segmentation}, which extends from the probabilistic U-Net~\cite{Kohl2018} to model prediction variation across multiple scales and generate multiple plausible predictions for each input. We extend the PHiSeg architecture with fully connected layers in the same way we extended the initial 3D U-Net architecture to regress cortical metrics predictions for each voxel.

We generate confidence maps for each prediction by sampling multiple times from each probabilistic network. This results in a range of segmentations and a range of metric predictions for each voxel. We argue that the larger the range of values predicted for a given voxel, the less confidence our network has in predicting a value for that voxel. We quantify the confidence of each prediction as the variance of each prediction made during inference. We seek the confidence map with the greatest correlation to prediction accuracy. From each range we can also define a metric 'Percentage in range' where we measure the percentage of voxels for which the ground truth value lies within the predicted range of values, where we seek the smallest ranges for which the ground truth is contained~\cite{Budd2019ConfidentSonographers}.

\section{Experimental Methods and Results}
\noindent\textbf{Data}: Data for this study comes from the Developing Human Connectome Project (dHCP) acquired in two stacks of 2D slices (in sagittal and axial planes) using a Turbo Spin Echo (TSE) sequence~\cite{Hughes2017ASystem}. The used parameters were: TR=12s, TE=156ms, SENSE factor 2.11 (axial) and 2.58 (sagittal) with overlapping slices. The study contains 505 subjects aged between 26-45 weeks. Tissue segmentations and surface metrics for training are derived form the dHCP surface extraction pipeline~\cite{Makropoulos2018TheReconstruction}\footnote{\href{https://github.com/BioMedIA/dhcp-structural-pipeline}{https://github.com/BioMedIA/dhcp-structural-pipeline}}.

A second clinical Down Syndrome cohort of 26 subjects was collected with a variety of scanning parameters, aged between 32-45 weeks, most subjects were acquired in the sagittal and transverse
planes using a multi-slice TSE sequence. Two stacks of 2D slices were acquired using the scanning parameters: TR=12s, TE=156ms, slice thickness = 1.6 mm with a slice overlap = 0.8 mm; flip angle = $90^{\circ}$ and an in-plane resolution: 0.8x0.8 mm.

\noindent\textbf{Preprocessing}: We project surface-based representations of cortical biometrics into a volumetric representation using a ribbon constrained method\footnote{\href{https://www.humanconnectome.org/software/workbench-command/-metric-to-volume-mapping}{https://www.humanconnectome.org/software/workbench-command/-metric-to-volume-mapping}} provided by the HCP project~\cite{Marcus2011InformaticsProject}. This operation is performed for both hemispheres of the brain and then combined into a single volume, where overlapping metric values are averaged. These volumes (together with the tissue segmentations) represent the training targets for the learning algorithm. T2w input volume and corresponding metric volumes are then resampled to a isotropic voxel spacing of 0.5 to ensure a prediction of physically meaningful values for each subject, \emph{i.e},  each voxel of our input image represents the same physical size in millimetres. Each T2 volume is intensity normalised to the range [0,1].

\noindent\textbf{Training}: 400 subjects are used for training; 50 for validation; 55 for test. During training we sample class-balanced 64x64x64 patches (N=12) from each subject's volume pair. We test on the entire volume of the image using a 3D sliding window approach. We conduct our experiments to predict two different cortical metrics: Thickness and Curvature.
\begin{figure}[h!]
\centering
    \begin{subfigure}[b]{.32\textwidth}
        \centering
        \includegraphics[width=\textwidth]{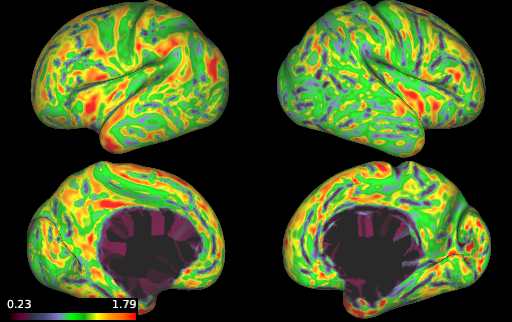}
        \caption{Ground truth thickness map}
        \label{fig:dhcp_t_gt}
    \end{subfigure}
\hfill
    \begin{subfigure}[b]{.32\textwidth}
        \centering
        \includegraphics[width=\textwidth]{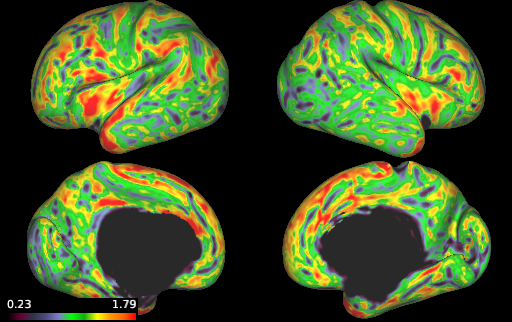}
        \caption{Predicted thickness map}
        \label{fig:dhcp_t_pred}
    \end{subfigure}
\hfill
    \begin{subfigure}[b]{.32\textwidth}
        \centering
        \includegraphics[width=\textwidth]{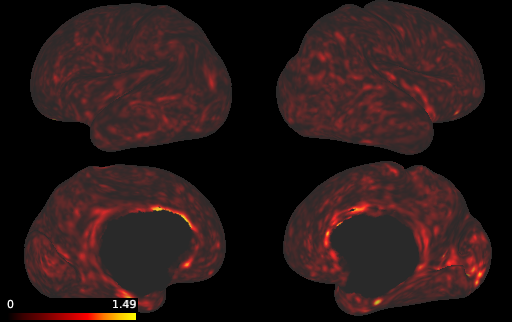}
        \caption{Thickness difference Map}
        \label{fig:dhcp_t_diff}
    \end{subfigure}
\vfill
    \begin{subfigure}[b]{.32\textwidth}
        \centering
        \includegraphics[width=\textwidth]{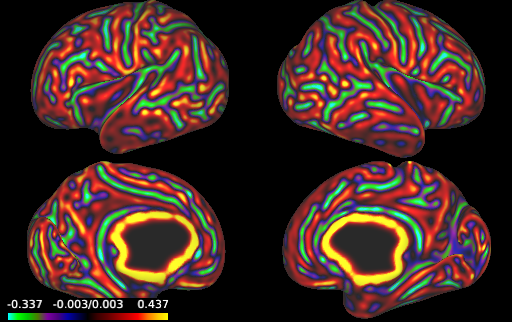}
        \caption{Ground truth curvature map}
        \label{fig:dhcp_c_gt}
    \end{subfigure}
\hfill
    \begin{subfigure}[b]{.32\textwidth}
        \centering
        \includegraphics[width=\textwidth]{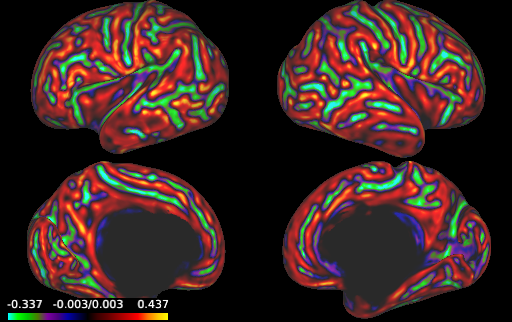}
        \caption{Predicted curvature map}
        \label{fig:dhcp_c_pred}
    \end{subfigure}
\hfill
    \begin{subfigure}[b]{.32\textwidth}
        \centering
        \includegraphics[width=\textwidth]{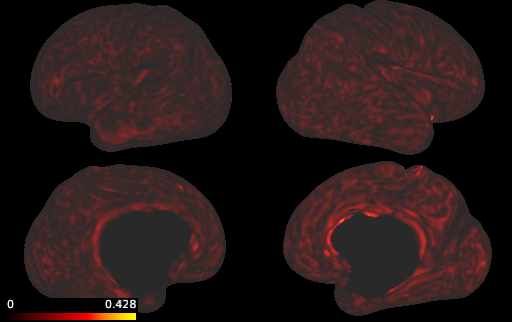}
        \caption{Curvature difference Map}
        \label{fig:dhcp_c_diff}
    \end{subfigure}
\caption{Qualitative results on a dHCP subject: Here we have used ground truth surfaces to re-project our predicted metric volumes and difference map back into a surface representation for ease of comparison.}\label{fig:dhcp_qual}
\end{figure}

\begin{table}[h!]
\centering
\begin{tabular}{llll}
\multicolumn{1}{l|}{\textit{\textbf{Experiment}}} & \multicolumn{1}{l|}{Dice} & \multicolumn{1}{l|}{Mean error (mm)} & Median error (mm)\\ \hline
Thickness (mm) &  &  &\\ \hline
\multicolumn{1}{l|}{\textbf{UNetMSE}} & \multicolumn{1}{l|}{\textbf{0.946 $\pm$ 0.010}} &  \multicolumn{1}{l|}{\textbf{0.179 $\pm$ 0.025}} &  \textbf{0.125 $\pm$ 0.021} \\
\multicolumn{1}{l|}{UNetHuber} & \multicolumn{1}{l|}{0.939 $\pm$ 0.012} & \multicolumn{1}{l|}{0.197 $\pm$ 0.027} & 0.139 $\pm$ 0.022 \\ \hline
Curvature &  &  &\\ \hline
\multicolumn{1}{l|}{\textbf{UNetMSE}} & \multicolumn{1}{l|}{\textbf{0.945 $\pm$ 0.010}} & \multicolumn{1}{l|}{\textbf{0.0424 $\pm$ 0.005}} & \textbf{0.0304 $\pm$ 0.003}  \\
\multicolumn{1}{l|}{UNetHuber} & \multicolumn{1}{l|}{0.935 $\pm$ 0.011} & \multicolumn{1}{l|}{0.0498 $\pm$ 0.006} & 0.0361 $\pm$ 0.004 
\end{tabular}
\caption{Results for metric prediction: Metric error calculated as average voxel-wise difference between prediction and ground truth only for voxels within the cortex. Dice score is reported over the segmentation.}\label{tab:det_thick}
\end{table}

\begin{figure}[h!]
\centering
    \begin{subfigure}[b]{.45\textwidth}
        \centering
        \includegraphics[width=\textwidth]{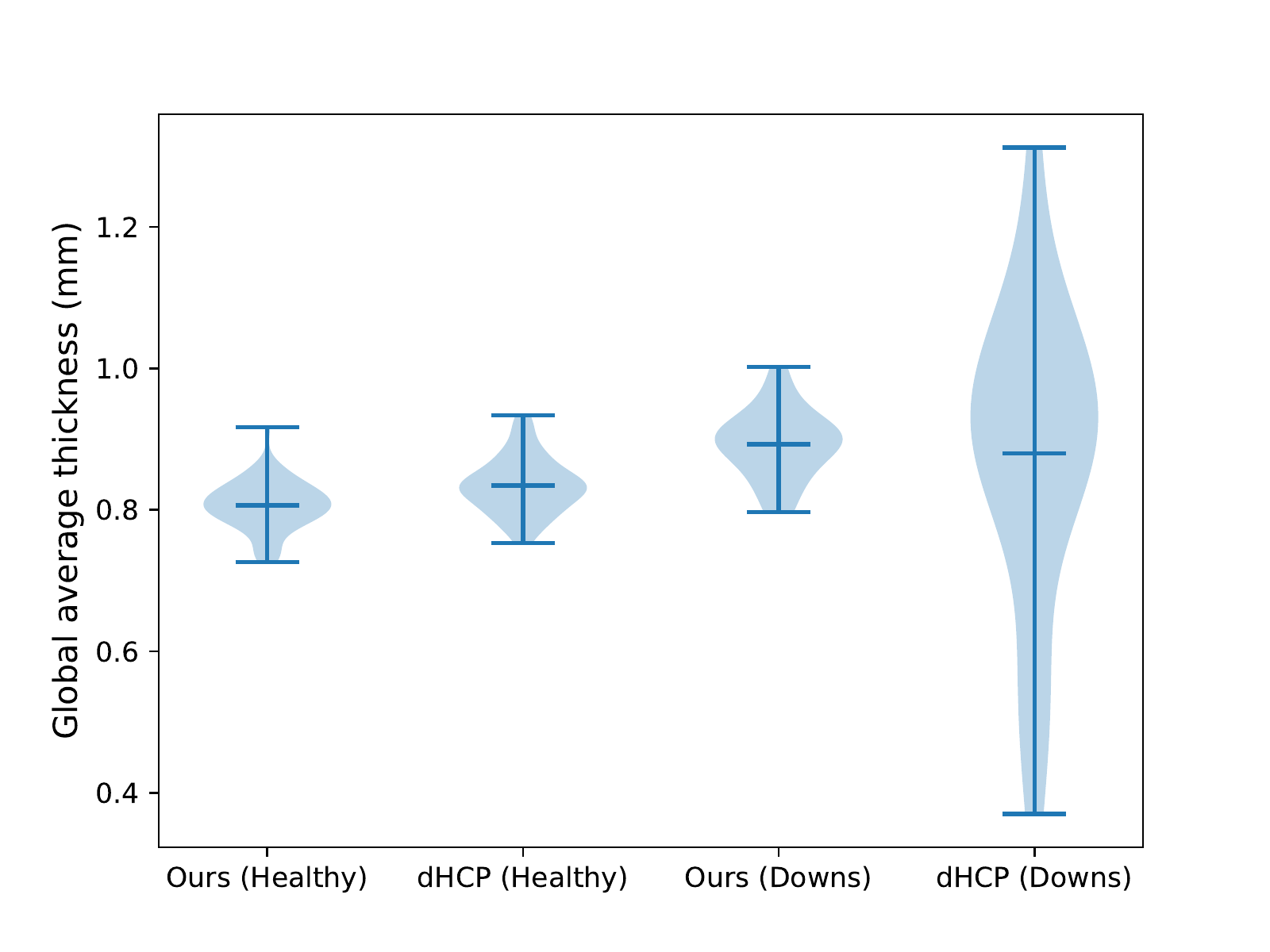}
        \caption{Thickness distributions}
        \label{fig:thick_vio}
    \end{subfigure}
    \begin{subfigure}[b]{.45\textwidth}
        \centering
        \includegraphics[width=\textwidth]{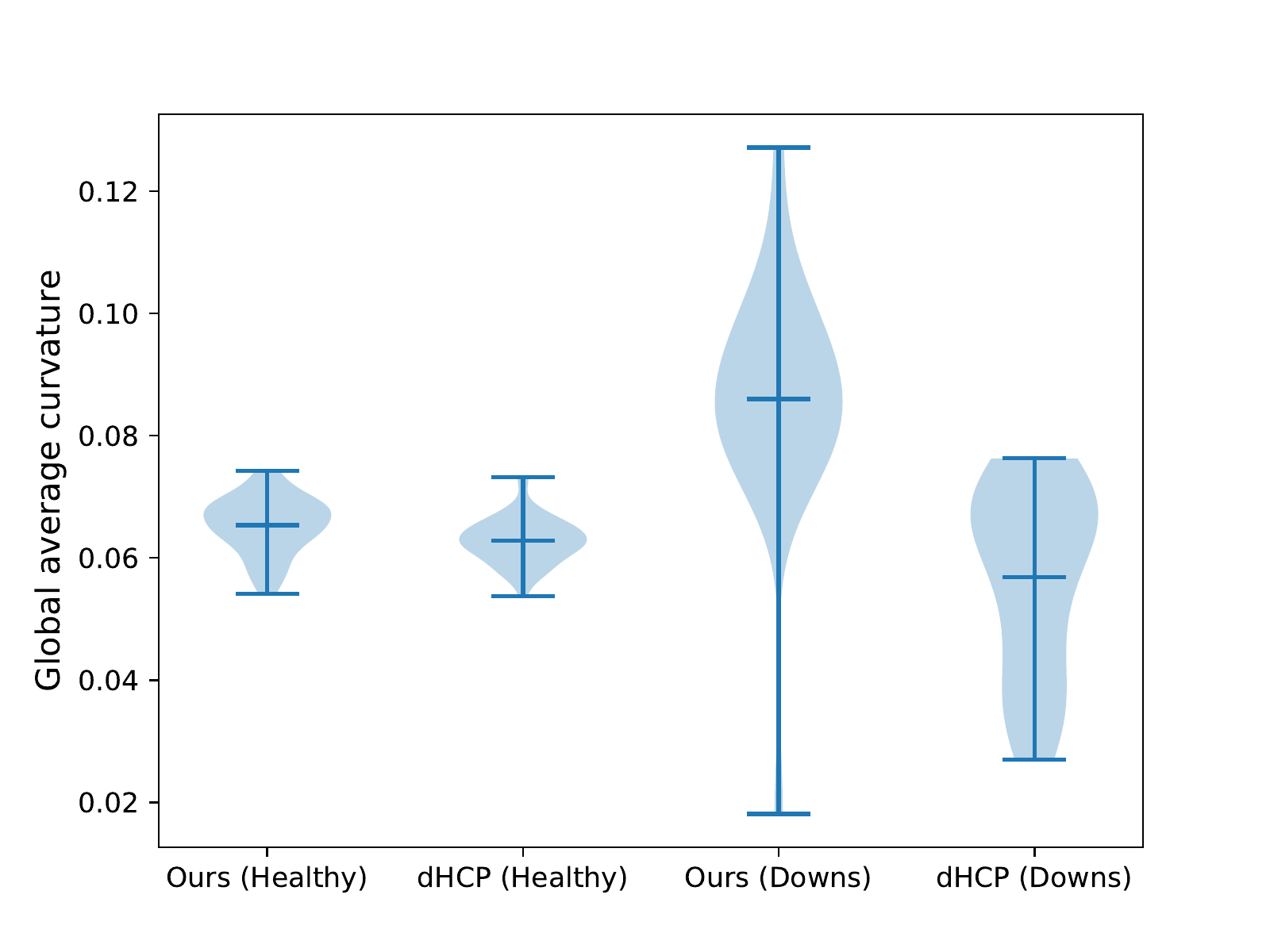}
        \caption{Curvature distributions}
        \label{fig:curv_vio}
    \end{subfigure}
    \caption{Global average metric distributions a) Thickness and b) Curvature. Our method produces a sensible distribution of values for the Down Syndrome data set compared to the dHCP pipeline for thickness prediction. However curvature prediction remains challenging. Pathological cases (right two in each plot) cannot be quantitatively evaluated because of missing ground truth.}\label{fig:downs_vio}
\end{figure}

\noindent\textbf{Deterministic experiments}: We establish a baseline for simultaneous estimation of cortical segmentation and metric regression. Table~\ref{tab:det_thick} reports performance measures for all deterministic experiments. We find minimal difference in performance using the Huber loss function instead of MSE loss on the regression task module. Figure \ref{fig:dhcp_qual} show example outputs produced by our best performing model in comparison to the ground truth. Our network successfully extracts accurate cortical metric predictions directly from the input MRI, maintaining the structural variation we expect across the cortex. We notice that some extreme values have not been accurately predicted, such as in cortical regions at the bridge between the two brain hemispheres (for curvature prediction). However the extreme values that are present in the ground truth data are an artifact of the surface based metric prediction method, hence it is less important to replicate this precisely. In Figure~\ref{fig:downs_vio} we report test-set wide metrics comparing predicted global metric distributions in comparison to ground truth distributions, our method predicts a similar distribution of results to the ground truth.

\begin{figure}
    \centering
    \includegraphics[width=0.8\textwidth]{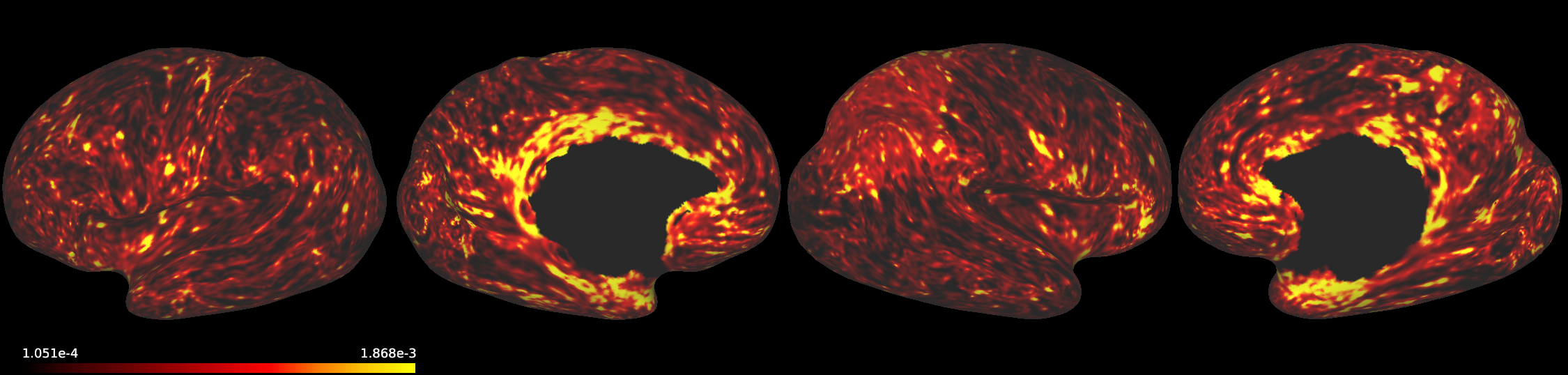}
    \caption{Confidence map generated with PHiSeg for a dHCP subject. Brighter areas indicate decreased confidence. }
    \label{fig:dhcp_conf}
\end{figure}

\noindent\textbf{Probabilistic experiments}: We consider probabilistic extensions of our previous best performing method. Table \ref{tab:prob_table} reports performance measures for all probabilistic experiments. We find that introducing DropBlock layers into our network has improved segmentation accuracy, but metric estimation accuracy has declined. 
PHiSeg has not improved either segmentation or metric estimation performance. In these experiments, PHiSeg training was often unstable, and took much longer to converge than other methods. 
While our error has increased, the ability to generate confidence maps for these predictions increases their value. Figure~\ref{fig:dhcp_conf} shows a generated confidence map using PHiSeg. 

\begin{table}[h!]
\centering
\begin{tabular}{lllll}
\multicolumn{1}{l|}{\textit{\textbf{Experiment}}} & \multicolumn{1}{l|}{Dice} & \multicolumn{1}{l|}{Mean Mean error} & \multicolumn{1}{l|}{Mean Median error} & Percent in range  \\ \hline
Thickness (mm) &  &  &\\ \hline
\multicolumn{1}{l|}{\textbf{UNetDropBlock}} & \multicolumn{1}{l|}{\textbf{0.945 $\pm$ 0.00}9} & \multicolumn{1}{l|}{\textbf{0.268 $\pm$ 0.017}} & \multicolumn{1}{l|}{\textbf{0.373 $\pm$ 0.017}} &  \textbf{59.83\%} \\
\multicolumn{1}{l|}{PHiSeg} & \multicolumn{1}{l|}{0.774 $\pm$ 0.075} & \multicolumn{1}{l|}{0.567 $\pm$ 0.096} & \multicolumn{1}{l|}{0.550 $\pm$ 0.052} & 19.60\%  \\ \hline
Curvature &  &  &\\ \hline
\multicolumn{1}{l|}{\textbf{UNetDropBlock}} & \multicolumn{1}{l|}{\textbf{0.946 $\pm$ 0.009}} & \multicolumn{1}{l|}{\textbf{0.0538 $\pm$ 0.004}} & \multicolumn{1}{l|}{\textbf{0.0705 $\pm$ 0.003}} & \textbf{56.19\%} \\
\multicolumn{1}{l|}{PHiSeg} & \multicolumn{1}{l|}{0.796 $\pm$ 0.022} & \multicolumn{1}{l|}{0.102 $\pm$ 0.006} &  \multicolumn{1}{l|}{0.102$\pm$ 0.006} & 24.98\%
\end{tabular}
\caption{Probabilistic Prediction results: We show dice scores and the mean metric error when taking mean and the median of multiple (N=5) predictions as the final output.}\label{tab:prob_table}
\end{table}

\begin{figure}[h!]
\centering
    \begin{subfigure}[b]{.49\textwidth}
        \centering
        \includegraphics[width=\textwidth]{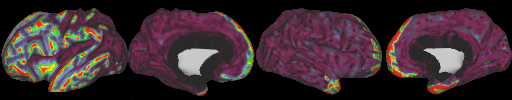}
        \caption{}
    \end{subfigure}
\hfill
    \begin{subfigure}[b]{.49\textwidth}
        \centering
        \includegraphics[width=\textwidth]{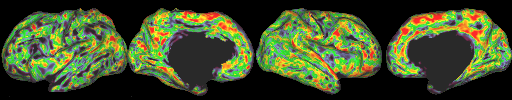}
        \caption{}
    \end{subfigure}
\vfill
    \begin{subfigure}[b]{.49\textwidth}
        \centering
        \includegraphics[width=\textwidth]{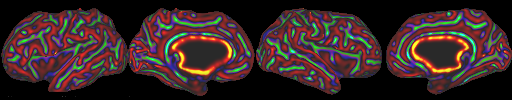}
        \caption{}
    \end{subfigure}
\hfill
    \begin{subfigure}[b]{.49\textwidth}
        \centering
        \includegraphics[width=\textwidth]{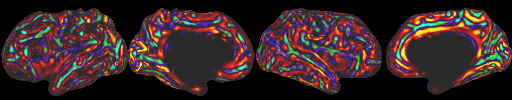}
        \caption{}
    \end{subfigure}
    \caption{Qualitative results on Down Syndrome dataset: a) dHCP predicted thickness; b) Our method predicted thickness; c) dHCP predicted curvature; d) Our method predicted curvature.}\label{fig:downs_surf}
\end{figure}

\noindent{\textbf{Down Syndrome experiments}}: We assess the performance of our method on a challenging Down Syndrome MRI dataset for which the dHCP pipeline fails to extract metrics correctly for many subjects. Since the 'ground truth' for this dataset is error prone, we demonstrate the performance of our method qualitatively and through population comparisons to the healthy dHCP test set used in previous experiments. Figure~\ref{fig:downs_surf} shows that our method produces more reasonable estimates of cortical thickness than the dHCP pipeline, thus showing evidence for our method's robustness to challenging datasets. Population statistics indicating that our method produces metric values in a sensible range for cortical thickness are shown in Figure~\ref{fig:downs_vio}. However our predicted distribution for cortical curvature is not consistent with healthy patients, indicating curvature and other more complex metrics remain challenging.

\section{Discussion}
We propose an automatic, pathology-robust method to predict cortical metric values from any T2w image stack. Our method is fast, robust and precise. 
We experiment with multi-task variants of the well known U-Net architecture and 
demonstrate how readily applicable deep learning is to predict cortical metric values in a reproducible way. 
Many architecture extensions are possible, which can for example be explored with neural architecture search methods~\cite{Elsken2018NeuralSurvey}.
In order to fully utilise the predictions of our network an imminent extension of our method is to automatically extract surface meshes from unseen data to enable proper visualisation of our predictions from images without ground truth surfaces. 

Probably the biggest advantage of our method is that we can produce cortical measurements for pathological cases. However, the curvature example in Fig.\ref{fig:curv_vio} shows that more complex metrics remain challenging and that we will need to further verify the robustness of our method in a clinical setting. Fine tuning may allow to generate disease-discriminative biomarkers directly from the network's latent space. 
At present our pipeline optimises cortical curvature and thickness prediction which naturally extends to sulcal depth and myelination prediction. There is potential to combine all metric predictions into a single model as it can be argued that prediction of different cortical properties are strongly correlated and would benefit each other as natural regulariser.

\section{Conclusion}
We offer an open-source framework for the extraction of cortical biometrics directly from T2 MRI images. This is the first of its kind that shows potential to be independent of age, image quality or presence of pathologies, and likely extendable to any cortical properties. We have tested our approach on a challenging pathological dataset for which we have not been able to reliably extract metrics with conventional methods like the dHCP processing pipeline. We expect that our future work will open new avenues for the analysis of cortical properties related to human brain development and disease in heterogenous populations.


\noindent\textbf{Acknowledgements:} 
We thank the parents and children who participated in this study.
This work was supported by the Medical Research Council [MR/K006355/1]; Rosetrees Trust [A1563], Fondation Jérôme Lejeune [2017b – 1707], Sparks and Great Ormond Street Hospital Children’s Charity [V5318]. The research leading to these results has received funding from the European Research Council under the European Unions Seventh Framework Programme (FP/2007–2013)/ERC Grant Agreement no. 319456. The work of E.C.R. was supported by the Academy of Medical Sciences/the British Heart Foundation/the Government Department of Business, Energy and Industrial Strategy/the Wellcome Trust Springboard Award [SBF003/1116].
We also gratefully acknowledge financial support from the Wellcome Trust IEH 102431, EPSRC (EP/S022104/1, EP/S013687/1), EPSRC Centre for Medical Engineering [WT 203148/Z/16/Z], the National Institute for Health Research (NIHR) Biomedical Research Centre (BRC) based at Guy's and St Thomas' NHS Foundation Trust and King's College London and supported by the NIHR Clinical Research Facility (CRF) at Guy’s and St Thomas’, and Nvidia GPU donations. 
%
%
\bibliographystyle{splncs04}
\bibliography{ref}
\newpage
\section*{Supplementary Material for 'Surface Agnostic Metrics for Cortical Volume Segmentation and Regression'}\label{app:A}

\begin{figure}
    \centering
    \begin{subfigure}{0.49\textwidth}
        \centering
        \includegraphics[width=\textwidth]{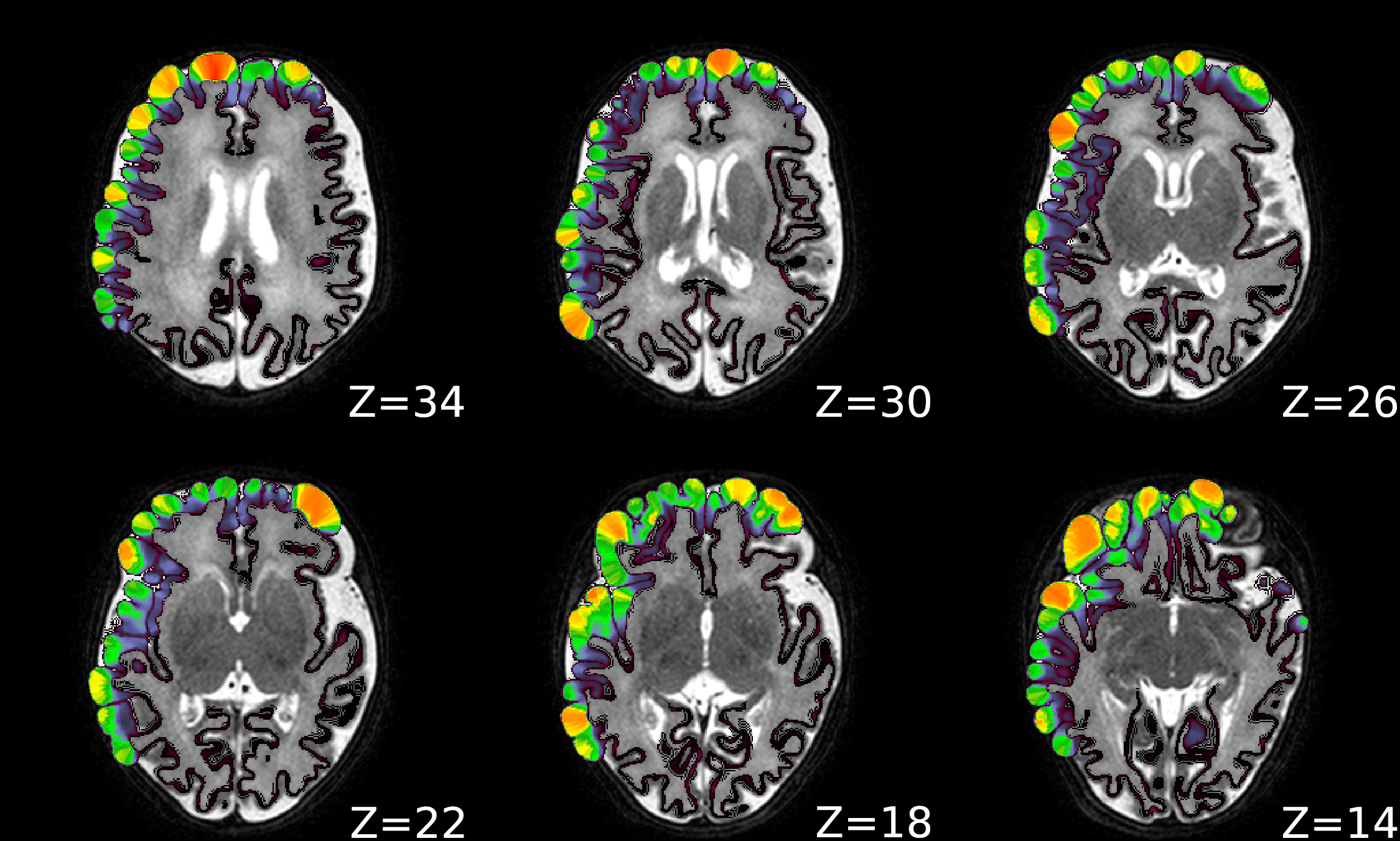}
        \caption{dHCP thickness prediction}
    \end{subfigure}
\hfill
    \begin{subfigure}{0.49\textwidth}
        \centering
        \includegraphics[width=\textwidth]{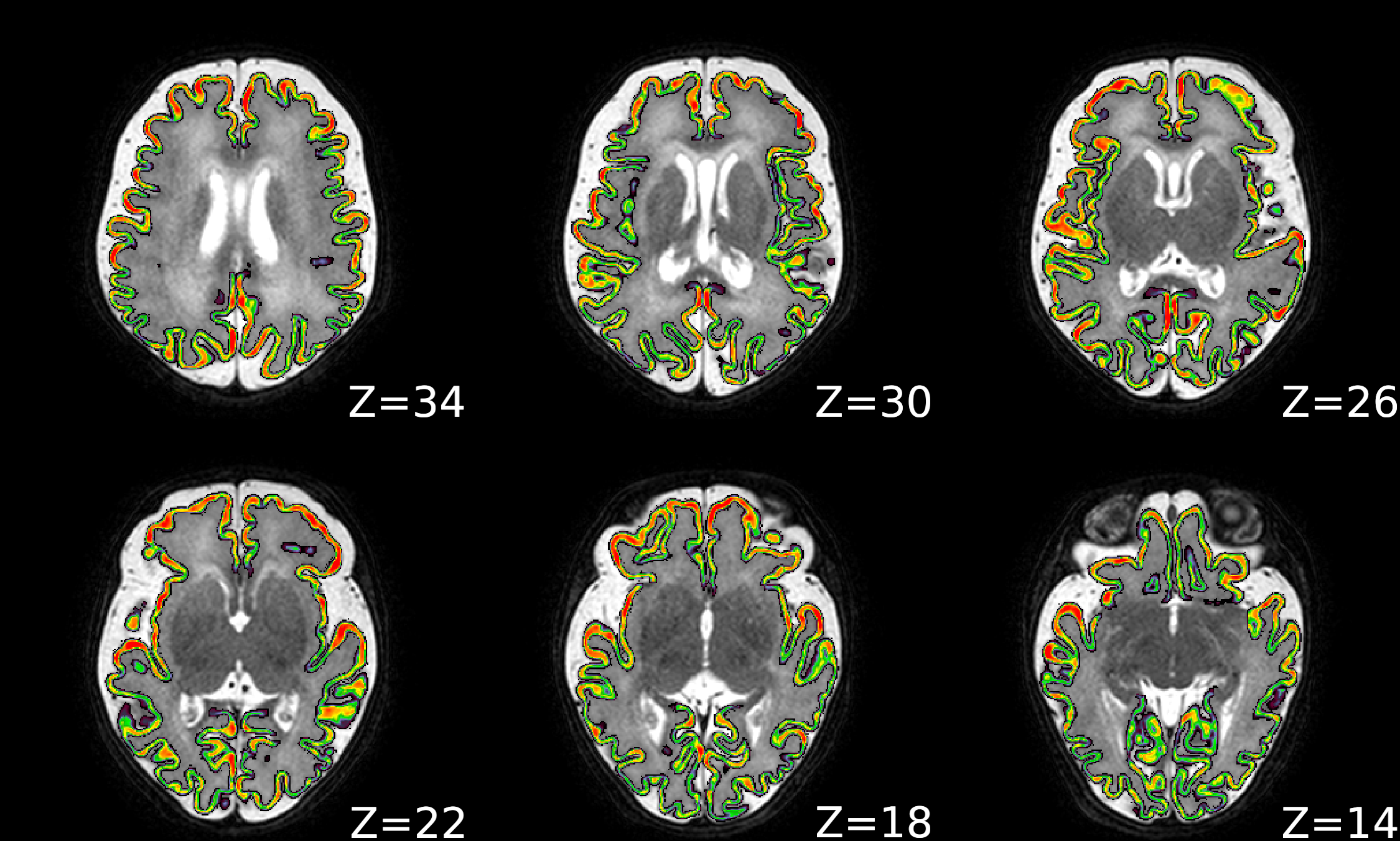}
        \caption{Our thickness prediction}
    \end{subfigure}
\vfill
    \begin{subfigure}{0.49\textwidth}
        \centering
        \includegraphics[width=\textwidth]{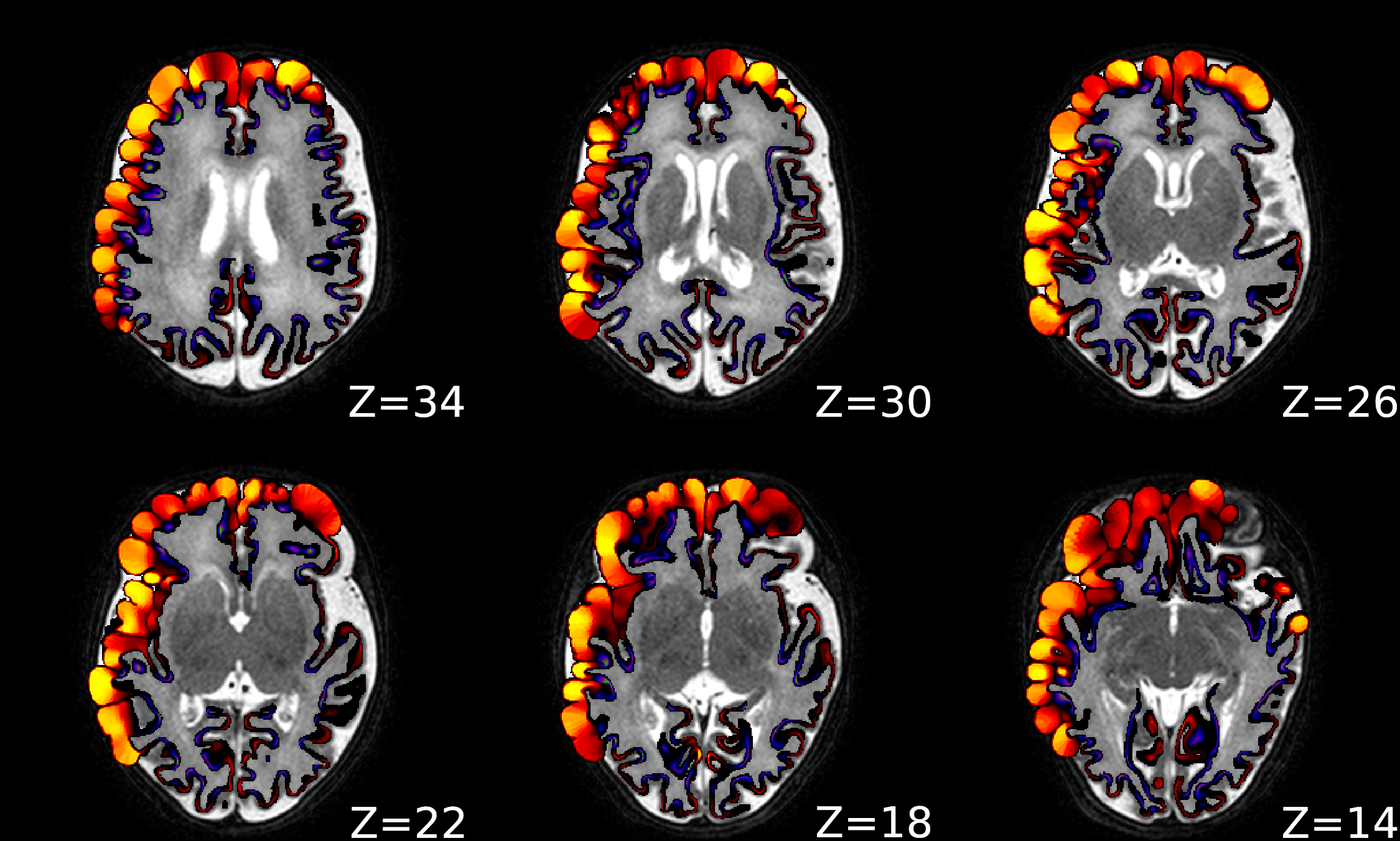}
        \caption{dHCP curvature prediction}
    \end{subfigure}
\hfill
    \begin{subfigure}{0.49\textwidth}
        \centering
        \includegraphics[width=\textwidth]{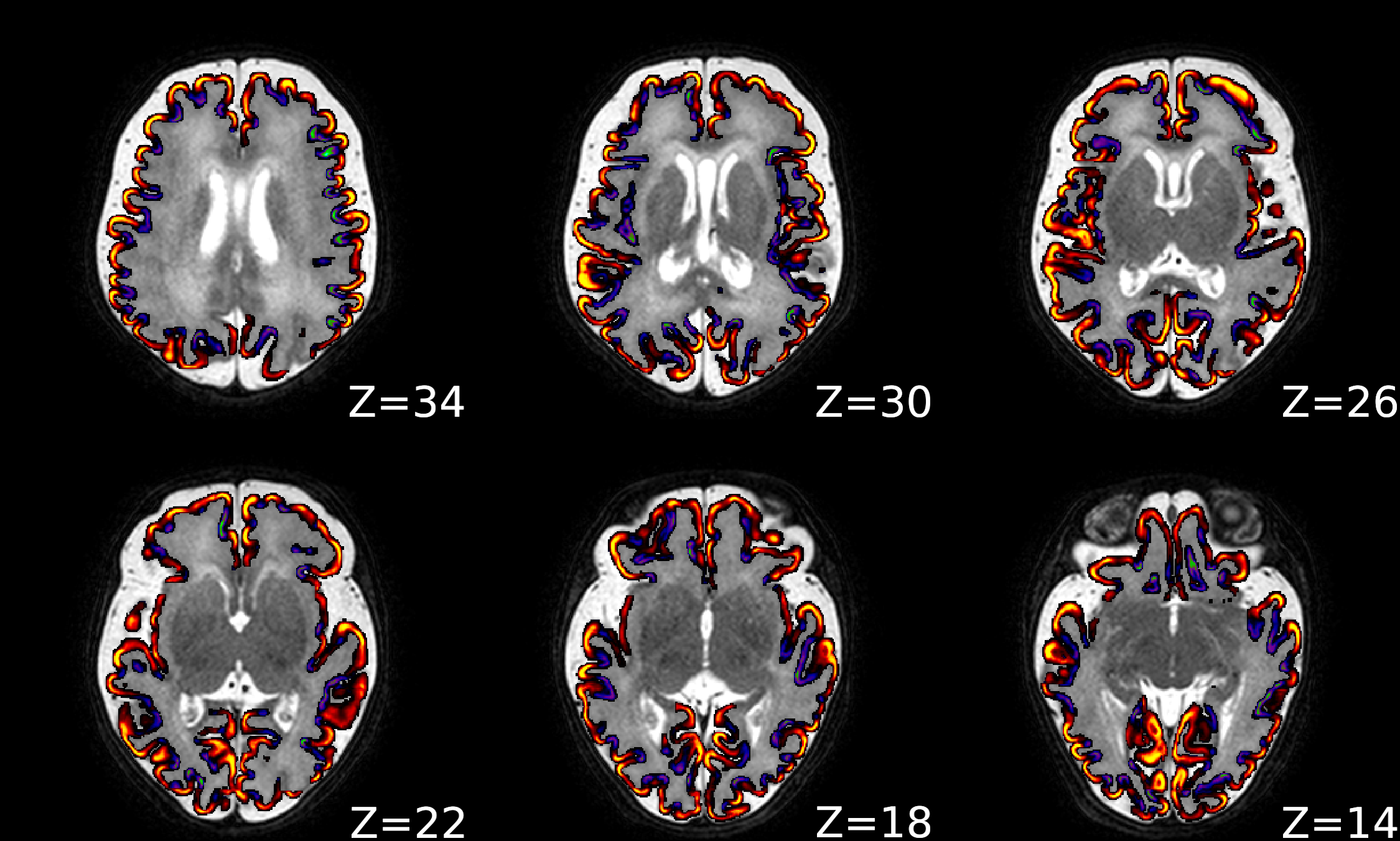}
        \caption{Our curvature prediction}
    \end{subfigure}
\vfill
    \caption{Slice view of our proposed method predictions. We can clearly see the artefacts generated using the dHCP pipeline for a Downs subject above, whereas our method produces a much more accurate segmentation and hence metric prediction.}
    \label{fig:slice_view}
\end{figure}

\end{document}